\documentclass[epj]{svjour}
\usepackage{graphics}
\begin{document}
\title{The Axial Form Factor of the Nucleon}
\author{Elizabeth Beise
}                     
\institute{University of Maryland, College Park, MD 20742 USA}
\date{Received: \today / Revised version: \today}
\abstract{The parity violation programs at MIT-Bates, Jefferson
Lab and Mainz are presently focused on developing a better
understanding of the sea-quark contributions to the vector matrix
elements of nucleon structure. The success of these programs will
allow precise semi-leptonic tests of the Standard Model such as
that planned by the QWeak collaboration. In order to determine
the vector matrix elements, a good understanding of the nucleon's
axial vector form factor as seen by an electron, $G^e_A$, is also
required. While the vector electroweak form factors provide
information about the nucleon's charge and magnetism, the axial
form factor is related to the nucleon's spin. Its $Q^2=0$ value
at leading order, $g_A$, is well known from nucleon and nuclear
beta decay, and its precise determination is of interest for
tests of CKM unitarity. Most information about its $Q^2$
dependence comes from quasielastic neutrino scattering and from
pion electroproduction, and a recent reanalysis of the neutrino
data have brought these two types of measurements into excellent
agreement. However, these experiments are not sensitive to
additional higher order corrections, such as nucleon anapole
contributions, that are present in parity-violating electron
scattering. In this talk I will attempt to review what is
presently known about the axial form factor and its various
pieces including the higher order contributions, discuss the the
various  experimental sectors, and give an update on its
determination through PV electron scattering.
} 
\maketitle
\section{Introduction}
\label{intro}

The neutral weak interaction between leptons and nucleons can be
described by a set of three form factors that contain information about
nucleon structure. The goal of present-day experiments in parity-violating
electron scattering experiments has been to determine the two vector weak
form factors, $G_E^Z$ and $G_M^Z$.  These two form factors can be used
along with the nucleon's electromagnetic form factors to disentangle the contributions
of up, down and strange quarks to the nucleon's charge and magnetization
distributions. However, in order to carry this out one also needs to
know the third weak form factor coming from the nucleon's
axial current, $G_A^Z$. The axial form factor
has been determined at low momentum transfer in
both quasielastic neutrino scattering and in pion
electroproduction, but very little information on $G_A^Z$ is available at
momentum transfers greater than 1~(GeV/c)$^2$. In addition, the axial
form factor as seen by an electron is substantially modified by electroweak
radiative corrections that cannot yet be computed with high precision,
including the term related to parity-violating coupling of a photon to
the nucleon, known as the anapole coupling.  Two experimental directions,
quasielastic neutrino scattering and parity-violating quasielastic electron
scattering from deuterium, can improve our knowledge of both $G_A^Z$ and
of the anapole contribution.

\section{Lepton-nucleon scattering}
\label{theory}

The nucleon electromagnetic current associated with
lepton-nucleon scattering can be written as

\begin{eqnarray*}
\langle N^{\prime}\vert J_\mu^\gamma \vert N \rangle &=&
\overline{u}_N \left[ F_1^\gamma(q^2)\gamma_\mu
+ \frac{i\sigma_{\mu\nu}q^\nu}{2M_N} F_2^\gamma(q^2)\right. \\
&+& \frac{G_F}{M_N^2}F_A^\gamma(q^2)
\left(q^2\gamma_\mu - q^\nu\gamma_\nu q_\mu\right)\gamma_5  \\
&-& \left. \frac{i\sigma_{\mu\nu}q^\nu\gamma_5}{2M_N}F_E(q^2)\right] u_N \, ,
\end{eqnarray*}
where $F_1^\gamma$ and $F_2^\gamma$ are the well-known Pauli and Dirac
electromagnetic form factors, $q^2=-Q^2$ is the four-momentum transfered from
the lepton to the nucleon. The term containing the anapole form
factor $F_A^\gamma(q^2)$ violates parity, and $F_E^\gamma(q^2)$ is a form factor
that would arise with time-reversal violation.  The anapole form factor has
been computed by several authors~\cite{Zhu00,Ris00,Mae00} and is expected
to be small at $Q^2=0$, but its computation is complicated by strong interaction
effects in the nucleon, and its momentum transfer dependence is unknown. It
is negligible in electron scattering cross section measurements but it
enters the asymmetry in parity-violating electron scattering at the same order as
the weak nucleon axial form factor. The nucleon's neutral weak current is
\begin{eqnarray*}
\langle N^{\prime}\vert J_{\mu}^Z + J_{\mu 5}^Z\vert N \rangle &=&
\overline{u}_N  \left[
F_1^Z(q^2)\gamma_\mu
+ \frac{i\sigma_{\mu\nu}q^\nu}{2M_N} F_2^Z(q^2)\right. \\
&+& \left. \gamma_\mu \gamma_5 G_A^Z(q^2)\right]u_N \, .
\end{eqnarray*}
The vector form factors $F_1^Z$ and $F_2^Z$ are of
primary interest in determining $s$-quark effects in PV electron
scattering, and the axial form factor $G_A^Z$ contains information about
the nucleon spin. At leading order, $G_A^Z$ can further be
explicitly deconstructed using SU(3) symmetry into isovector and
isospin singlet components to separate out the contribution of $s$-quarks to nucleon spin
\begin{displaymath}
G_A^Z(Q^2) = -\tau_3 G_A(Q^2) + G_A^s(Q^2) \, ,
\end{displaymath}
where $\tau_3 = +1(-1)$ for $p(n)$, $G_A(0)=-(g_A/g_V)=1.2670\pm 0.0035$~\cite{PDG04}
as determined in nucleon $\beta$ decay, and $G_A^s(0)=\Delta s$, the strange quark
spin content of the nucleon. The $Q^2$ dependence of $G_A$ has generally been
characterized by a dipole form, $1/(1+Q^2/M_A^2)^2$, which can then be linked
to a determination of an axial radius in a low momentum expansion of $G_A$ with $Q^2$:
\begin{displaymath}
\langle r_A^2\rangle = -\frac{6}{g_a} \frac{dG_A}{dQ^2}\vert_{Q^2=0} = \frac{12}{M_A} \, .
\end{displaymath}

\section{Available Data}
\label{data}

Two methods have been used to determine this lowest order $Q^2$ behavior of
the axial form factor. The most direct method is to use quasielastic
neutrino-nucleon scattering. Very little neutral current scattering data is available,
so cross section data from the charged current process
$\nu_\mu + n \rightarrow \mu^- + p$ has typically been use to extract $M_A$. Recently,
a new global fit to neutrino data was carried out by Budd, {\it et al.}
~\cite{Bud03}, which improved over earlier fits both by using the most
recent determination of $(g_A/g_V)$ along with new results for nucleon electromagnetic
form factors.  The improved fit gives $M_A = 1.001\pm 0.020$~GeV.  In pion
electroproduction, $M_A$ can also be extracted from the transverse component of the
near-threshold electroproduction cross section by associating it with the electric
dipole transition amplitude $E^{(-)}_{0+}$ through the low energy theorem of
Nambu, Lurie and Shrauner~\cite{Nam62} under the assumption that $m_\pi = 0$. A measurement
was recently carried out at the Mainz Microtron~\cite{Lie99}, resulting in
$M_A = 1.068\pm 0.015$~GeV. In a recent topical review, Bernard~{\it et al.}~\cite{Ber02}
used chiral
perturbation theory to compute a finite mass correction to this extraction, which
is substantial and results in a corrected $M_A$ of $1.013\pm 0.015$~GeV, bringing
it into agreement with the neutrino data.  Therefore, it
appears that $M_A$ is reasonably well determined and the low $Q^2$ behavior of $G_A$
can at least be described phenomenologically. This does not, however, give a
first principles theoretical description of $G_A(Q^2)$, nor does it provide an
adequate description at momentum transfers above 1~(GeV/c)$^2$.
In addition, better modeling of neutrino scattering, guided by improved data, will
be required for upcoming neutrino oscillation experiments. A new
experiment, Miner$\nu$a~\cite{McF04}, has been proposed that would consist of a
high granularity neutrino detector located at the NUMI beam line at Fermilab. This
experiment would be able to provide a precise determination of $G_A$, including
possible departures from the nominal dipole behavior,
at $Q^2<2$~(GeV/c)$^2$ and a first determination of $G_A$ for $Q^2>2$~(GeV/c)$^2$.
Planning for another potential experiment is underway at Jefferson Lab, using
the reaction $\vec{e}+p\rightarrow \nu + n$, covering the range
$Q^2\sim 1-3$~(GeV/c)$^2$~\cite{Deu04}. This very challenging experiment would require
detection of the recoiling neutrons at very forward angles, and the parity-violating
asymmetry in the detected neutrons would
be measured in order to constrain backgrounds.

While the above measurements would be able to better constrain $G_A(Q^2)$ and
provide improved models of cross section data for neutrino oscillation experiments,
it is the axial form factor as seen by an electron that is relevant to the parity
violation program that is the topic of this workshop. The axial form factor seen
in PV electron scattering can be written, going beyond first order, as
\begin{displaymath}
G_A^e(Q^2) = -\tau_3(1+R_A^{T=1})G_A(Q^2) + R_A^{T=0}G_A^8(Q^2) + G_A^s(Q^2)
\end{displaymath}
where $R_A^{T=0,1}$ are electroweak radiative corrections arising from
higher order diagrams~\footnote{I am here following the notation in~\cite{Mus94}.}.
The SU(3) octet form factor $G_A^8$ is not present
at tree-level, but appears once radiative corrections are included. Its
$Q^2=0$ value can be estimated from the ratio of axial vector to vector couplings
in hyperon $\beta$ decay which, assuming SU(3) flavor symmetry, can be related
to the octet axial charge $a_8$ and to the hyperon $F$ and $D$
coefficients~\cite{PDG04},
\begin{displaymath}
G_A^8(0) = \frac{(3F-D)}{2\sqrt{3}} = \frac{1}{2}a_8 = 0.217\pm 0.043\, .
\end{displaymath}
Its $Q^2$ behavior has also not been measured, but it is usually
assumed to have the same dipole form  as the isovector form factor
$G_A(Q^2)$ with the same mass parameter $M_A$.

It should be noted that while a decade of measurements related to
the ``spin crisis'' have indirectly determined $G_A^s(0)=\Delta s$ from
polarized deep-inelastic scattering, its $Q^2$ behavior is also unknown. There has
been one determination of $G_A^s(0)$ from quasielastic neutrino scattering~\cite{Ahr87},
which is in reasonable agreement with the polarized DIS data. An improved analysis
of these data was carried out by Garvey~{\it et al.}~\cite{Gar93} who included
possible effects of nonzero strange vector form factors. A recent further improved
analysis was carried out by S.~Pate~\cite{Pat04}, who combined the neutrino data with
results from HAPPEX to perform a global fit to the three strange form factors
(so far with only two constraints) to extract $G_A^s$ at the mean of the two
experiments, $Q^2=0.5$~(GeV/c)$^2$, rather than extrapolating to $Q^2=0$.
A new direct measurement of $G_A^s$ at low momentum transfer
has been proposed using the ratio of neutral current to charged current neutrino
scattering at low momentum transfer,
FiNeSSE, using a highly segmented detector with wavelength shifting optical
fibers embedded in mineral oil to identify tracks left by
the recoiling protons~\cite{Tay04}. This experiment would potentially improve the
determination of $\Delta s$ by about a factor of two over
the DIS data, and with less theoretical uncertainty.

Of more direct interest to the parity violation program is the radiative correction to
the isovector $G_A(Q^2)$, of which the nucleon's anapole form factor $F_A^\gamma$
is one component. The dominant contributions to $R_A^{T=0}$ and $R_A^{T=1}$ come
from 1-quark terms such as $\gamma$-$Z$ mixing and vertex corrections, which
have been computed by several authors~\cite{Zhu00,PDG04}. Multi-quark or anapole
contributions were also computed~\cite{Zhu00}, who modeled them in terms of
hadronic parity-violating NN couplings cast within a heavy baryon chiral
perturbation theory framework. The results are shown in Table~\ref{tab:1}.
While the 1-quark contributions dominate the correction, the anapole contributions
dominate the uncertainty.

\begin{figure}
\resizebox{0.5\textwidth}{!}{\includegraphics{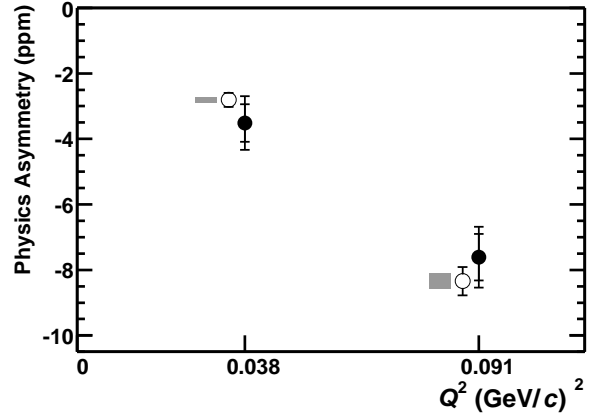}}
\caption{Asymmetry results from the two SAMPLE deuterium experiments
(solid circles, see~\protect{\cite{Ito04}}),
compared to expectation from theory using the axial radiative corrections
of~\protect{\cite{Zhu00}} (open circles). The theory also assumes a value of $G_M^s$ of
0.15 nuclear magnetons, and the grey band represents a change in $G_M^s$ of
$\pm$0.6 n.m.}
\label{fig:1}       
\end{figure}

\begin{figure}
\resizebox{0.5\textwidth}{!}{\includegraphics{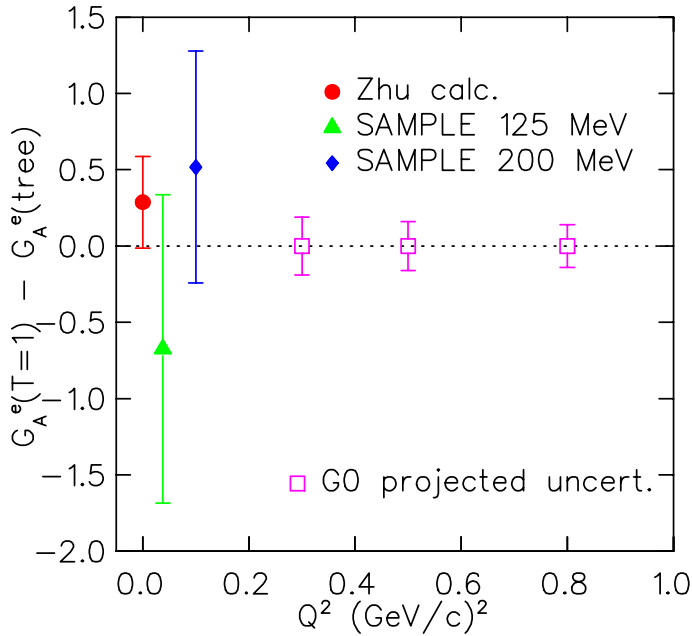}}
\caption{The difference between the axial form factor seen by
an electron in parity-violating electron scattering and $G_A^Z(Q^2)$, the
tree-level form factor. Shown are the calculation of~\protect{\cite{Zhu00}},
SAMPLE data, and projected uncertainties in the G0 experiment. }
\label{fig:2}       
\end{figure}

\begin{table}
\caption{Electroweak radiative corrections, computed in the $\overline{MS}$ scheme,
for the axial form factor measured in PV electron scattering. The values are
taken from~\protect{\cite{Zhu00}}.}
\label{tab:1}       
\begin{tabular}{lll}
\hline\noalign{\smallskip}
Source & $R_A^{T=1}$ & $R_A^{T=0}$  \\
\noalign{\smallskip}\hline\noalign{\smallskip}
1-quark & $-0.18$ &  $0.07$ \\
anapole & $-0.06\pm 0.24$ & $0.01\pm 0.14$ \\
total   & $-0.24\pm 0.24$ & $0.08\pm 0.14$ \\
\noalign{\smallskip}\hline
\end{tabular}
\end{table}
The axial form factor $G_A^e$, or at least its isovector piece ${G_A^e}^{(T=1)}$,
can be determined from the PV asymmetry in quasielastic scattering from deuterium,
where the strange quark effects in the neutron and proton tend to cancel.  Nuclear
effects, including both parity conserving~\cite{Dia01} and
parity-violating~\cite{Sch03,Liu03} contributions, have been shown to be small.  The
first measurement of ${G_A^e}^{(T=1)}$ was carried out by the SAMPLE collaboration.
The measured asymmetry at two momentum transfers are shown in Figure~\ref{fig:1}.
They agree fairly well with the calculation, which was carried
out at $Q^2=0$, indicating that there is no anomalously large $Q^2$ dependence
to the anapole term or to the correction at these very low momentum
transfers. However, very little else is known about
its behavior away from $Q^2=0$.  Two model calculations~\cite{Mae00,Ris00}
of $F_A^\gamma(Q^2)$ have been carried out, and they indicate a much
softer behavior with $Q^2$ than that of $G_A^Z(Q^2)$, even possibly
an increase with $Q^2$, as well as quite different behavior for the isoscalar
and isovector pieces. These could substantially
enhance the effects of radiative corrections at momentum transfers in the range of
the G0 experiment. It would thus be very useful to have some experimental
information on ${G_A^e}^{(T=1)}$ at higher momentum transfers. A program
of backward angle measurements with a deuterium target is part of the planned
running for the G0 experiment, and aerogel {\v Cerenkov} detectors have been
added to the detector array in order to identify and separate charged pions
produced in the deuterium target from the desired quasielastically scattered
electrons.  These data will not only reduce the model uncertainties in the
determination of $G_E^s$ and $G_M^s$ from the hydrogen data, but will also
allow the first experimental information on $G_A^e$ away from the static limit.
Shown in Figure~\ref{fig:2} are the projected uncertainties from the
G0 experiment~\cite{G0-04}
in the difference between ${G_A^e}^{(T=1)}$ and the tree-level $G_A^Z$ that is
seen in neutrino scattering, along with the calculation of~\cite{Zhu00}
and the two SAMPLE measurements.

%

As an aside, it should be noted that, assuming that a determination of the nucleon axial form
factor can straightforwardly be related to electron-quark interactions, the two SAMPLE
measurements can be recast in terms of the two electron-quark couplings $C_{2u}$
and $C_{2d}$.  Prior to the SAMPLE measurements, experimental limits on these were
from the original SLAC DIS parity-violation experiment~\cite{Pre79}, and from
the parity-violating quasielastic elecron scattering experiment on $^9$Be carried out at
Mainz~\cite{Hei89}.  The two SAMPLE measurements are sensitive to the combination
$C_{2u}-C_{2d}$.  These are modified by 1-quark radiative corrections, and in the
case of elastic $e$-$N$ scattering the multiquark corrections as well.
In order to compare directly to the SLAC DIS data,
the multi-quark radiative corrections must be removed, which
although small, dominate the uncertainty. The resulting values from the 200~MeV
and 125~MeV data sets, respectively, are
\begin{eqnarray*}
C_{2u}-C_{2d} &=& -0.042 \pm 0.040 \pm 0.035 \pm 0.02 \\
C_{2u}-C_{2d} &=& -0.12 \pm 0.05 \pm 0.05 \pm 0.02 \pm 0.01 \, ,
\end{eqnarray*}
where the first two uncertainties are statistical and experimental systematic,
the third is that due the radiative corrections, and, for the 125~MeV case,
the last corresponds to variations in $G_M^s$ by $\pm$0.6 because it is undetermined
at this momentum transfer.  These values are in good agreement with the
Standard Model prediction~\cite{PDG04}, and represent a significant improvement over the
earlier data.

\section{Conclusion}
\label{summary}

In summary, while much attention has been focused on determination of the
neutral weak vector form factors, in order to extract strange quark effects
in the nucleon, there are variety of experimental avenues to pursue in the near
future to improve our knowledge of the nucleon's axial form factor. The tree-level
form factor is now known reasonably well at low momentum transfers from neutron
beta decay, from quasielastic neutrino scattering and from pion electroproduction.
Its knowledge at higher momentum transfer, including potential deviations from
a generic dipole behavior, can be improved with new neutrino scattering experiments, and these
experiments will help provide the required precision cross section information needed
for the next generation of neutrino oscillation measurements.  In PV electron scattering,
the axial form factor is substantially modified and very little is known about the
$Q^2$ behavior of the higher order terms. The G0 experiment will uniquely be able
to provide the higher $Q^2$ data through quasielastic scattering from a deuterium target.

%
%

%

\end{document}